\documentclass[sigconf]{acmart}
\setcopyright{none}
\renewcommand\footnotetextcopyrightpermission[1]{}
\AtBeginDocument{%
  }

\setcopyright{acmlicensed}
\copyrightyear{2025}
\acmYear{2025}
\acmDOI{XXXXXXX.XXXXXXX}

\usepackage{algorithmic}
\usepackage{makecell}
\usepackage{graphicx}
\usepackage{xcolor}
\usepackage{hyperref}
\usepackage{caption}
\usepackage{subcaption}
\usepackage{textcomp}
\usepackage{xcolor}

\newcommand{\pcdiff}[1]{\textcolor{red}{\emph{{#1}}}}

\begin{document}

\title{Is RISC-V ready for High Performance Computing? An evaluation of the Sophon SG2044}

\author{Nick Brown}
\email{n.brown@epcc.ed.ac.uk}
\orcid{0000-0003-2925-7275}
\affiliation{%
  \institution{EPCC at the University of Edinburgh}
  \city{Edinburgh}  
  \country{UK}
}


\begin{abstract}
The pace of RISC-V adoption continues to grow rapidly, yet for the successes enjoyed in areas such as embedded computing, RISC-V is yet to gain ubiquity in High Performance Computing (HPC). The Sophon SG2044 is SOPHGO's next generation 64-core high performance CPU that has been designed for workstation and server grade workloads. Building upon the SG2042, subsystems that were a bottleneck in the previous generation have been upgraded.

In this paper we undertake the first performance study of the SG2044 for HPC. Comparing against the SG2042 and other architectures, we find that the SG2044 is most advantageous when running at higher core counts, delivering up to 4.91 greater performance than the SG2042 over 64-cores. Two of the most important upgrades in the SG2044 are support for RVV v1.0 and an enhanced memory subsystem. This results in the SG2044 significantly closing the performance gap with other architectures, especially for compute-bound workloads. 
\end{abstract}

\keywords{RISC-V, High Performance Computing (HPC), Sophon SG2044, NAS Parallel Benchmark suite}

\maketitle
\pagestyle{plain}
\section{Introduction}
The RISC-V Instruction Set Architecture (ISA) has garnered significant attention in recent years within the computing community due to its open nature, modular design, and extensibility. As a community driven ISA, RISC-V provides a flexible foundation for hardware vendors to develop innovative processor designs without the constraints imposed by proprietary standards. Its growing adoption in recent years across diverse computing domains, from embedded microcontrollers to automotive, underscores its potential as a competitive alternative to established ISAs such as x86 and Arm.

Indeed, in 2024 Nvidia announced that RISC-V cores are used as management controllers in every one of their GPUs, and indeed they estimated that they had shipped over a billion RISC-V cores in 2024 alone \cite{rv-nvidia}. Embedded controllers aside, recent advancements in RISC-V-based hardware has further accelerated interest in its role as a main line processor, or accelerator, for High Performance Computing (HPC) workloads. 

SOPHGO's 64-core Sophon SG2042 was the first high core count commodity available RISC-V CPU that had potential for HPC. Since it's release in 2023, it has been widely explored across numerous areas including computational fluid dynamics \cite{brown2023performance}, astrophysics \cite{diehl2024preparing}, and as a RISC-V build server. However, \cite{brown2023performance} found that whilst the SG2042 delivered fairly impressive compute performance, limitations in the memory subsystem meant that it struggled to perform for workloads that involved anything beyond simple memory access patterns.

\begin{table*}[htb]
    \centering
    \caption{Summary of memory behaviour for NPB benchmark kernels on a Xeon Platinum 8170 from \cite{brown2023performance}.}
    \label{tab:npb_description}
    \begin{tabular}{|c|c|c|c|}
    \hline
      \textbf{Benchmark}   & \makecell{\textbf{Clock ticks} \\ \textbf{cache stall}} & \makecell{\textbf{Clock ticks} \\ \textbf{DDR stall}} & \makecell{\textbf{Time DDR} \\ \textbf{bandwidth bound}} \\
     \hline
	  \textbf{Integer Sort (IS)} & 35\% & 0\% & 16\% \\        
	  \textbf{Multi Grid (MG)} & 34\% & 20\% & 88\% \\
        \textbf{Embarrassingly Parallel (EP)} & 11\% & 0\% & 0\% \\
        \textbf{Conjugate Gradient (CG)} & 19\% & 18\% & 0\% \\
        \textbf{Fast Fourier Transform (FT)} & 13\% & 9\% & 18\% \\
        \hline
        \textbf{Block Tridiagonal (BT)} & 8\% & 9\% & 0\% \\
        \textbf{LU Gauss Seidel (LU)} & 12\% & 11\% & 0\% \\
        \textbf{Scalar Pentadiagonal (SP)} & 20\% & 21\% & 0\% \\
        
    \hline
    \end{tabular}
\end{table*}

When SOPHOGO announced the Sophon SG2044 they promised it would have significant improvements in memory performance along with version 1.0 of the RISC-V Vector (RVV) standard. Consequently the key question explored in this paper is, given the ability to drive the vector unit with mainline compilers and promised memory performance to keep this fed with data, does the SG2044 move RISC-V towards being ready for use in HPC? 

This paper presents a performance evaluation of SOPHGO's next generation CPU, the Sophon SG2044, which has been designed for high-throughput computing. Using NASA's Parallel Benchmark suite (NPB), we compare and contrast performance delivered by the SG2044 against other RISC-V CPUs, including the SG2042, along with those more commonly found in HPC that leverage other ISAs.

This paper is structured as follows; Section \ref{sec:bg} explores the background to this work by describing the SG2044 in more detail and highlighting lessons learnt from the SG2042, along with describing the benchmark suite used to evaluate this processor. Section \ref{sec:other_rv_comparison} then compares a single C920v2 core in the SG2044 against single cores of other popular, commodity available, RISC-V CPUs before Section \ref{sec:sg2042-compare} compares against the SG2042. In Section \ref{sec:otherarch} we compare multi-core scaling performance of the SG2044 against other CPUs which are popular in HPC, before highlighting performance differences between compiler versions and the benefits of vectorisation in Section \ref{sec:compiler}. Conclusions from this work are drawn in Section \ref{sec:conclusions} which also discussed future avenues for performance analysis of the SG2044.

\section{Background}
\label{sec:bg}
\subsection{The Sophon SG2044}

The SG2044 builds upon its predecessor, the 64-core SG2042, by upgrading specific subsystems. Retaining the underlying approach of providing a 64-core processor organised in clusters of four T-Head XuanTie cores, the SG2042 contains the C920v2 core. Similarly to the C920v1 in the SG2042, the C920v2 provides a 128-bit wide vector unit, however this now conforms to version 1.0 of the RISC-V Vector extension (RVV) standard unlike the C920v1 which only followed RVV v0.7.1. This lack of following the ratified standard meant that mainline compilers were unable to target vectorisation for the SG2042 as these only support RVV v1.0. Consequently, a potential benefit of the SG2044 is that developers are able to vectorise their code using mainline GCC and LLVM, rather than having to use a bespoke fork of the compiler that implements RVV v0.7.1 and potentially lags the latest compiler enhancements. SOPHGO have not published the clock frequency for the SG2044 and whilst \cite{sg2044-config} suggests that this is 2.8 GHz, in the test system used for this work it is 2.6 GHz. By contrast, the SG2042 runs at 2 GHz.

One of the most important enhancements in the SG2044 for HPC is the improvement in memory performance. Indeed, \cite{brown2025investigations} identified that this was one of the primary bottlenecks found in the SG2042, and \cite{sg2044-announce} announced that the SG2044 would deliver around a threefold increase in DDR memory bandwidth. Unlike the SG2042, SOPHGO engineers have informed us that all cores in the SG2044 are within a single NUMA region and there are 32 memory controllers and 32 memory channels, compared with 4 memory controllers and 4 memory channels within the SG2042.

Each C920v2 core is 64-bit with a 12 stage out-of-order multiple issue superscalar pipeline design. Providing the RV64GCV instruction set, the C920v2 has three decode, four rename/dispatch, eight issue/execute and two load/- store execution units. Each C920v2 core contains 64KB of L1 instruction (I) and data (D) cache. Furthermore, there is 2MB of L2 cache shared between each cluster of four cores \cite{sg2044-config}, which is double that of the SG2042, and 64MB of L3 cache shared by all cores in the processor.

Compared to the SG2042, the SG2044 also provides upgraded I/O capabilities which includes compatibility with PCIe Gen5, compared to Gen4 in the SG2042, and support for DDR5 memory. The SG2044 can therefore be considered a logical progression of the SG2042, where performance bottlenecks that were identified in the SG2042 have been addressed and subsystems upgraded to provide compatibility with latest standards.

In mid-2025 support for the SG2044 was integrated into the Linux 6.16 mainline kernel, providing upstream Linux support. Preliminary Geekbench performance results are somewhat confusing, \cite{sg2044-evb-geekbench} reports impressive performance against the SG2042 however these tests use different versions of the benchmark suite, Geekbench 6.4.0 on the SG2044 and Geekbench 6.3.0 on the SG2042. By contrast, \cite{sg2044-geekbench} compared these two CPUs using the same, Geekbench 6.3.0, and performance between the CPUs was similar for single core workloads and the SG2044 was on average around 1.3 times faster for multi-core benchmarks \cite{sg2044-geekbench}. Irrespective, from the individual Geekbench results it is clear that, compared to the SG2042, the SG2044 excels for some multi-core workloads such as \emph{Horizon Detection} which is four times faster and \emph{File Compression} which is two times faster \cite{sg2044-geekbench}.

\begin{table*}[htb]
    \centering
    \caption{Single core comparison between RISC-V technologies with performance reported in Mops/s (Higher is better) using NPB kernels running at class B. In red is the percentage performance delivered compared to the C920v2 core of the SG2044.}
    \label{tab:risc-v-core-comparison}
    \begin{tabular}{|c|c|c|c|c|c|c|c|}
    \hline
      \textbf{Benchmark}   & \textbf{SG2044} & \textbf{VisionFive V2} & \textbf{VisionFive V1} & \textbf{SiFive U740} & \textbf{All Winner D1} & \textbf{Banana Pi} & \textbf{Milk-V Jupyter}\\
     \hline
	IS & 64.68 & \makecell{17.84 \\ \pcdiff{(28\%)}} & \makecell{6.36 \\ \pcdiff{(10\%)}} & \makecell{9.09 \\ \pcdiff{(14\%)}} & \makecell{5.41  \\ \pcdiff{(8\%)}} & \makecell{22.66  \\ \pcdiff{(35\%)}} & \makecell{24.75  \\ \pcdiff{(38\%)}}\\    
	MG & 1472.32 & \makecell{288.65 \\ \pcdiff{(20\%)}} & \makecell{72.31 \\ \pcdiff{(5\%)}} & \makecell{90.28 \\ \pcdiff{(6\%)}} & \makecell{163.19 \\ \pcdiff{(11\%)}} & \makecell{306.78  \\ \pcdiff{(22\%)}} & \makecell{335.38  \\ \pcdiff{(23\%)}}\\    
    EP & 40.75 & \makecell{12.01 \\ \pcdiff{(30\%)}} & \makecell{7.55 \\ \pcdiff{(19\%)}} & \makecell{9.08 \\ \pcdiff{(22\%)}} & \makecell{9.23 \\ \pcdiff{(23\%)}} & \makecell{18.17 \\ \pcdiff{(45\%)}} & \makecell{20.4 \\ \pcdiff{(50\%)}}\\
    CG & 269.37 & \makecell{43.61 \\ \pcdiff{(16\%)}} & \makecell{21.96 \\ \pcdiff{(8\%)}} & \makecell{29.09 \\ \pcdiff{(11\%)}} & \makecell{12.99 \\ \pcdiff{(5\%)}} & \makecell{23.71 \\ \pcdiff{(9\%)}} & \makecell{24.42 \\ \pcdiff{(9\%)}}\\         
    FT & 1296.22 & \makecell{245.99 \\ \pcdiff{(19\%)}} & \makecell{88.35 \\ \pcdiff{(7\%)}} & \makecell{116.59 \\ \pcdiff{(9\%)}} & DNR & \makecell{362.8 \\ \pcdiff{(28\%)}} & \makecell{388.24 \\ \pcdiff{(30\%)}}\\    
    \hline
    \end{tabular}
\end{table*}

\subsection{NASA's NAS Parallel Benchmark suite}

The NAS Parallel Benchmarks (NPB) were first developed in 1991 by NASA’s Numerical Aerodynamic Simulation (NAS) program, now the advanced supercomputing division, to evaluate supercomputer performance. The suite was designed to approximate real-world scientific workloads that are of interest to NASA, whilst being architecturally agnostic. The original suite, which we focus on here, comprises eight benchmarks which include five kernels and three pseudo applications. These eight benchmarks mimic key algorithmic patterns that are typically found in Computational Fluid Dynamics (CFD) as well as other high-performance codes. All benchmarks are configured using a variety of problem sizes, known as classes, and the NAS team have developed a range of implementations, including the OpenMP version used in this paper. Throughout our experiments we use the benchmark code without any modifications.

Table \ref{tab:npb_description} provides a summary of memory behaviour for each benchmark in the NAS Parallel Benchmark suite when executed with OpenMP across all 26 physical cores of an Intel Xeon Platinum 8170 processor from \cite{brown2023performance}. The columns \emph{Clock ticks cache stall} and \emph{Clock ticks DDR stall} indicate the proportion of CPU cycles spent stalled due to cache and main memory accesses, respectively. The \emph{Time DDR bandwidth bound} column shows the percentage of total execution time during which DDR bandwidth utilisation was high.

The IS kernel, which exercises indirect and random memory access patterns, causes significant cache-related stalls, highlighting its irregular access behaviour. The MG kernel is notably memory-bound, exhibiting high stall rates at both cache and DDR levels, as well as sustained periods of high DDR bandwidth utilisation. In contrast, EP has been designed to test compute performance with minimal memory pressure and this results in negligible memory stalls and no periods of elevated DDR utilisation.

The CG kernel, which involves irregular memory accesses combined with nearest-neighbour communications, spends approximately 37\% of clock cycles stalled on memory operations. FT, which requires all-to-all communication for parallel data transposition, exhibits a more moderate 22\% stall rate, but sustains high DDR bandwidth utilisation for 18\% of the execution time which is more than any other kernel apart from MG.

The pseudo-application benchmarks BT, LU, and SP are more complex than the kernels and represent real-world HPC simulations. All three solve the 3D compressible Navier–Stokes equations using finite-difference methods. LU employs a block-lower block-upper triangular approximation based on a Gauss–Seidel iterative scheme \cite{saphir1997new}. BT and SP solve the same system using a Beam–Warming approximation, where BT leverages a block-tridiagonal system and SP fully diagonalises the equations. Both these codes use Gaussian elimination to solve the system of equations. Among the three, BT exhibits the lowest memory access stall rates, while SP incurs the highest, as reflected in Table \ref{tab:npb_description}.

\section{Comparison against other RISC-V CPUs}
\label{sec:other_rv_comparison}
In this section we compare performance of the SG2044 against existing commodity RISC-V solutions. Due to the difference in core counts between RISC-V CPUs, we focus here on single core performance to understand how the XuanTie C920v2 core of the Sophon SG2044 performs against other widely available RISC-V cores. We compare against the U74 core \cite{u74} which is contained in the JH7200 and JH7100 SoCs of the VisionFive V2 and V1 respectively, both of these boards contain 8GB of DRAM. We also compare against the SiFive Freedom U740 SoC, also containing the U74 core and 16GB of DDR, the T-Head XuanTie C906 \cite{c906} in the AllWinner D1 SoC with 1GB of memory, the SpacemiT K1 which is in the Banana Pi BPI-F3 \cite{bananapi}, and SpacemiT M1 in the Milk-V Jupyter. The SpacemiT K1/M1 in the Banana Pi and Milk-V Jupyter is especially interesting because it is the only other core in this group, apart from the SG2044, that implements RVV v1.0, it also provides 256-bit vector width and is RVA22 compliant \cite{spacemit}.

Table \ref{tab:risc-v-core-comparison} presents a single-core performance comparison across these RISC-V CPUs using the five NAS Parallel Benchmark (NPB) kernels at problem size class B. GCC version 15.2, which is the latest release version of GCC at the time of writing, is used throughout these experiments, and vectorisation is enabled for the SG2044, Banana Pi and Milk-V Jupyter (apart from for the CG benchmark, see Section \ref{sec:compiler}). Performance is measured in millions of operations per second (Mop/s), where higher values indicate better performance. For reference, the percentage performance that is achieved relative to a single T-Head C920v2 core (as used in the SG2044) is shown in italicised red.

Across all benchmarks, the SG2044's C920v2 consistently delivers the highest performance, significantly outperforming all other evaluated RISC-V cores. Among the alternatives, with the exception of the CG benchmark, the SpacemiT K1/M1 in the Banana Pi and Milk-V Jupyter achieves the closest performance to the C920v2, although only once reaches half the performance of the C920v2. The Milk-V Jupyter marginally outperforms the Banana Pi BPI-F3 for all benchmarks and this is likely because of minor differences between the Banana Pi BPI-F3's SpacemiT K1 and the SpacemiT M1 in the Milk-V Jupyter. Whilst both of these CPUs are based on the SpacemiT X60 RISC-V core, the SpacemiT M1 is a higher-clocked, better cooled, version of the SpacemiT K1 and-so the Milk-V Jupyter runs at 1.8 GHz whereas the Banana Pi BPI-F3 at 1.6 GHz. 

SiFive's U74 core in the VisionFive V2 achieves the next, although significantly lower, performance compared to the C920v2. While the VisionFive V1 and SiFive HiFive Unmatched (U740) also feature the U74 core, their performance is substantially lower which is consistent with prior findings in \cite{brown2023risc}. The T-Head C906 core, found in the Allwinner D1 SoC, lags behind both the C920v2 and VisionFive V2’s U74 in most benchmarks with the exception of the EP and MG kernels, where it outperforms the VisionFive V1 and SiFive U740. Due to the limited, 1GB, memory on the Allwinner D1 it was not possible to run the FT benchmark.

Overall, these results demonstrate that the SG2044's C920v2 core delivers significantly superior single-core performance compared to other commodity available RISC-V CPUs. Whilst this aligns with prior observations in \cite{brown2023risc} and \cite{brown2023performance}, the comparison here has been extended to include the SG2044's C920v2, along with the SpacemiT K1/M1. 

\section{Comparison against the SG2042}
\label{sec:sg2042-compare}
Section \ref{sec:other_rv_comparison} demonstrated that a single C920v2 core delivers impressive performance compared to other popular, commodity available, RISC-V CPUs. However, many of these other CPUs were designed primarily for use in embedded computing, and-so a key question is how the SG2044 compares against the SG2042 which is also a CPU designed for server grade workloads.

Table \ref{tab:sg_singlecore_comparison} provides a single-core performance comparison between the SG2044 and SG2042 for the NAS kernels running at class C. We experimented with different versions of the compiler, including GCC v15.2, for the SG2042 and found that T-Head's fork of GCC v8.4 (XuanTie GCC) consistently provided the best performance and-so it is that compiler version used for the SG2042 in both this section and Section \ref{sec:otherarch}. GCC v15.2 was used for the SG2044. From Table \ref{tab:sg_singlecore_comparison} it can be seen that, whilst a single C920v2 core in the SG2044 consistently outperforms a single SG2042 C920v1 core, the difference is fairly marginal. Indeed it is the compute bound, EP, kernel at 30\% where the SG2044 enjoys the greatest performance benefit over the SG2042. It should be highlighted that these results demonstrate a greater relative benefit for the SG2044 than the Geekbench 6.3 performance results in \cite{sg2044-geekbench}.

\begin{table}[htb]
    \centering
    \caption{Performance of NAS kernels (class C) running on a single C920v2 core of the SG2044 agsinst the C920v1 of the SG2042.}
    \label{tab:sg_singlecore_comparison}
    \begin{tabular}{|c|c|c|c|}
    \hline
      \textbf{Benchmark}   & \makecell{\textbf{SG2044} \\ \textbf{\textit{Mop/s}}} & \makecell{\textbf{SG2042} \\ \textbf{\textit{Mop/s}}} & \makecell{\textbf{SG2044} \\ \textbf{times faster}} \\
     \hline
	  \textbf{IS} & 63.63 & 58.87 & 1.08 \\
	  \textbf{MG} & 1382.91 & 1175.69 & 1.18 \\
        \textbf{EP} & 40.76 & 31.36 & 1.30 \\
        \textbf{CG} & 213.82 & 173.39 & 1.23 \\
        \textbf{FT} & 1023.83 & 797.09 & 1.28 \\
        
    \hline
    \end{tabular}
\end{table}

Given the memory limitations of the SG2042 \cite{brown2023performance}, it might seem surprising that the performance difference between the SG2044 and SG2042 is not greater for the IS (memory latency bound) and MG (memory bandwidth bound) kernels. Indeed it is EP, the compute bound kernel, that delivers the greatest relative benefit here and this is potentially due to the C920v2 providing RVV v1.0 along with the memory performance to keep the vector unit fed.

\begin{table}[htb]
    \centering
    \caption{Performance of NAS kernels (class C) running on all 64-cores of the SG2044 and SG2042 using OpenMP.}
    \label{tab:sg_multicore_comparison}
    \begin{tabular}{|c|c|c|c|}
    \hline
      \textbf{Benchmark}   & \makecell{\textbf{SG2044} \\ \textbf{\textit{Mop/s}}} & \makecell{\textbf{SG2042} \\ \textbf{\textit{Mop/s}}} & \makecell{\textbf{SG2044} \\ \textbf{times faster}} \\
     \hline
	  \textbf{IS} & 3038.14 & 618.50 & 4.91 \\        
	  \textbf{MG} & 32457.83 & 14397.69 & 2.25 \\
        \textbf{EP} & 2538.38 & 1675.25 & 1.52 \\
        \textbf{CG} & 7728.80 & 3508.95 & 2.20 \\
        \textbf{FT} & 22582.2 & 8317.91 & 2.71 \\        
    \hline
    \end{tabular}
\end{table}

A single core has the entirety of the L2 cache which is shared between clusters of four cores, the 64MB L3 cache shared between all cores, and all the off-chip memory bandwidth. Therefore when exploring the difference in memory performance between the SG2044 and SG2042 it is more instructive to run over all 64-cores and stress the memory subsystem. Table \ref{tab:sg_multicore_comparison} reports results from this multi-core performance comparison, threading across the 64-cores with OpenMP. For this experiment it can be seen that the SG2044 significantly outperforms the SG2042 for all kernels. 

\begin{table*}[htb]
    \centering
    \caption{Summary of CPUs that are benchmarked in this section.}
    \label{tab:other_arch}
    \begin{tabular}{|c|c|c|c|c|c|c|}
    \hline
      \textbf{CPU} & \textbf{ISA} & \textbf{Part} & \textbf{Base clock} & \makecell{\textbf{Number} \\ \textbf{of cores}} & \textbf{Vector} \\
     \hline
	 AMD EPYC & x86-64 & EPYC 7742 & 2.25GHz & 64 & AVX2\\
      Intel Skylake & x86-64 & Xeon Platinum 8170 & 2.1 GHz & 26 & AVX512\\      
      Marvell ThunderX2 & ARMv8.1 & CN9980 & 2 GHz & 32 & NEON\\
      Sophon SG2042 & RV64GCV & SG2042 & 2 GHz & 64 & RVV v0.7.1\\
      Sophon SG2044 & RV64GCV & SG2044 & 2.6 GHz & 64 & RVV v1.0.0\\
    \hline
    \end{tabular}
\end{table*}

By contrast to the single core comparison, when running across all 64-cores the compute bound kernel, EP, benefits the least from the SG2044, whereas the IS benchmark which tests memory latency performance, benefits the most. This demonstrates that SOPHGO have, as they claimed in \cite{sg2044-announce}, significantly improved the memory performance of the SG2044 compared to the SG2042 and this becomes most apparently when running workloads across all the cores and there is contention for cache and external memory access. The CG and FT kernels also include significant amounts of memory access and in \cite{brown2023performance} this limited their performance on the SG2042. By contrast, these kernels are both over two times faster when run across 64-cores of the SG2044.

\begin{figure}[htb]
    \centering
    {\includegraphics[width=1.0\columnwidth]{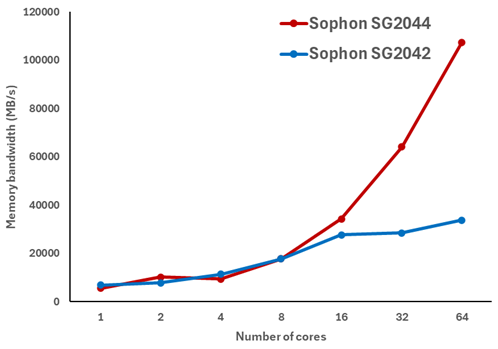}}
    \caption{Memory bandwidth reported by STREAM benchmark (higher is better) when run with different numbers of cores.}
    \label{fig:stream}
\end{figure}

To understand the memory performance differences between the SG2044 and SG2042 in more detail we ran the STREAM benchmark which is a synthetic benchmark to measures sustainable memory bandwidth. Figure \ref{fig:stream} illustrates the memory bandwidth reported for the\emph{copy} kernel, which only involves data movement and no computation, on both CPUs for different numbers of cores. It can be seen that memory performance of an individual core in the SG2044 is comparable to a core in the SG2042, and indeed up to and including 8 cores the memory bandwidth provided by both CPUs is very similar. However, beyond this point memory bandwidth of the SG2044 continues to scale with an increasing number of cores, whereas the SG2042 plateaus. Indeed, at 64 cores the SG2044 delivers over three times higher memory bandwidth than the SG2042 and this corresponds to what SOPHOGO promised in \cite{sg2044-announce}.

\section{Comparison against other architectures}
\label{sec:otherarch}

To understand whether the SG2044 is a contender for HPC workloads it is instructive to benchmark against CPUs that implement other ISAs and are commonly used for HPC. An overview of the selected processors is provided in Table \ref{tab:other_arch}.

We compare against the AMD EPYC 7742, part of AMD's Rome series, which is based on the Zen 2 microarchitecture. Benchmarking of this CPU is undertaken on ARCHER2, the UK national supercomputing service, which is a Cray EX. The EPYC 7742 features 64 physical cores distributed across four NUMA regions (16 cores per region) and includes eight memory controllers and eight memory channels. Each core integrates 32 KB of L1 instruction and data cache, 512 KB of L2 cache, and shares 16 MB of L3 cache per 4-core group. Supporting AVX2, the EPYC has 256-bit vector registers, double the width of the SG2044’s RVV units, and can execute two AVX-256 operations per cycle. The ARCHER2 nodes used in this study are equipped with 256 GB of DDR4 memory, and all code was compiled using GCC version 11.2 which is the latest version on the machine. Simultaneous multithreading (SMT) was disabled, in accordance with ARCHER2’s default configuration.

The Intel Xeon Platinum 8170 is a Skylake-SP processor and this was also used for profiling in Table \ref{tab:npb_description}. The Skylake comprises 26 physical cores, each with 32 KB of L1 instruction and data cache, 1 MB of L2 cache, and a shared 35.75 MB L3 cache (approximately 1.375 MB per core). The Skylake supports AVX-512, providing 512-bit wide vector operations, which are two and four wider than the vector units of the EPYC 7742 and SG2042, respectively. Each core contains two floating-point units (FPUs). The system used has 192 GB of DDR4 memory, and benchmarks were compiled with GCC version 8.4.

We also include results for the Marvell ThunderX2 CN9980, an ARMv8.1 (AArch64) processor based on the Vulcan microarchitecture. This processor features 32 physical cores, each with 32 KB of L1 instruction and data cache, 256 KB of private L2 cache, and 32 MB of shared L3 cache (1 MB per core). The ThunderX2 supports NEON SIMD, with 128-bit vector registers—equivalent in width to the vector units of the SG2044’s C920 cores. Like the Skylake, each core integrates two FPUs. This CPU is deployed in Fulhame, an HPE Apollo 70 system, with 128 GB of DDR4 memory per node. Benchmarks on this system were compiled with GCC version 9.2, and SMT was disabled during all runs.

For the benchmarks conducted in this section we execute class C of the NAS Parallel Benchmark (NPB) suite using the OpenMP implementations \cite{jin1999openmp}. Each thread is explicitly bound to a distinct physical core to avoid interference from thread migration or hyper-threading. All results represent the average of five independent runs, and binaries were compiled with -O3 optimization level.

\subsection{Integer Sort (IS)}

\begin{figure}[htb]
    \centering
    {\includegraphics[width=1.0\columnwidth]{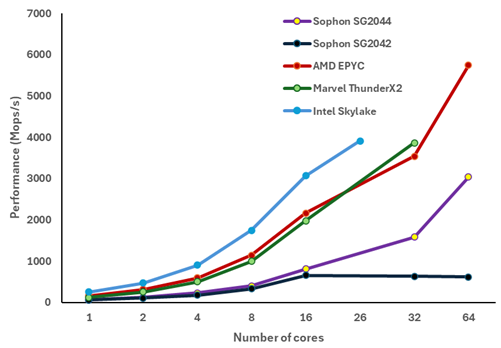}}
    \caption{IS benchmark performance (higher is better) parallelised via OpenMP.}
    \label{fig:is}
\end{figure}

The Integer Sort (IS) benchmark is memory latency bound and undertakes indirect, random, memory accesses to compare integers. Figure \ref{fig:is} illustrates performance results for this benchmarks across our CPUs of interested, reported in Mops/s (higher is better). In \cite{brown2023performance} it was observed that the SG2042 performs considerably worse than all other CPUs for this benchmark, with performance on the SG2042 plateauing at 16 cores. Performance on the SG2044 is fairly comparable to the SG2042 up to and including 8 threads, but then continues scaling with the number of cores, following a similar curve, albeit lower in absolute terms, to the AMD CPU. This re-enforces conclusions from the memory bandwidth results reported in Figure \ref{fig:stream}, where the ability for the SG2044 to continue scaling beyond 8 threads is ultimately what delivers the 4.91 times better performance at 64 threads compared to the SG2042.

It can be seen that even though the SG2044 delivers considerably better performance than the SG2042 for this benchmark kernel, the RISC-V CPUs still lags the other architectures. When running on a single core, the AMD EPYC delivers around twice the performance of the SG2044 and the Intel Skylake around three times. Therefore whilst the SG2044 scales well for this workload, at the individual core level there is likely still some overhead for this memory latency bound workloads compared to these other architectures.

\subsection{Multi Grid (MG)}

\begin{figure}[htb]
    \centering
    {\includegraphics[width=1.0\columnwidth]{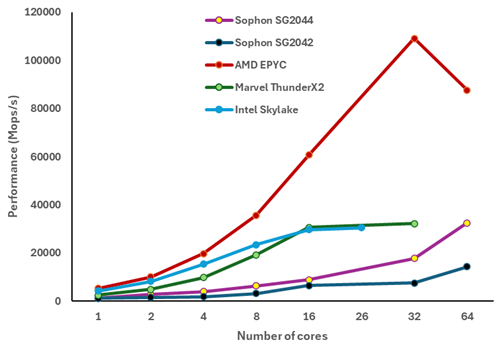}}
    \caption{MG benchmark performance (higher is better) parallelised via OpenMP.}
    \label{fig:mg}
\end{figure}

The Multi Grid (MG) benchmark is memory bandwidth bound, and results of running this benchmark kernel on the CPUs of interest is illustrated in Figure \ref{fig:mg}. It can be seen that, similarly to the IS kernel, as we scale the number of cores the difference between the SG2044 and SG2042 becomes most apparent. When comparing equal numbers of cores between architectures the SG2044 lags behind the AMD, Intel and Arm CPUs. However when considering entire CPU performance, running on all cores 26 cores of the Intel and 32 cores of the Arm CPUs, the SG2044 is comparable to those two other architectures whereas the SG2042 falls behind considerably.

The CPU's memory configuration partially explains the performance observed in Figure \ref{fig:mg}, where the AMD EPYC has 8 memory controllers and 8 memory channels, connected to DDR4-3200 memory. The Skylake and ThunderX2 both only have 2 memory controllers and are both connected to DDR4-2666, with the ThunderX2 having 8 memory channels compared to 6 memory channels in the Skylake. 

The SG2042 has four memory controllers and four memory channels connected to DDR4-3200, whereas the SG2044 has 32 memory controllers and 32 memory channels connected to DDR5-4266. This increase in memory controllers and memory channels, along with the use of DDR5, is a major part of the enhancement SOPHGO have made to the memory subsystem in the SG2044. When running over 64 cores the ratio of cores to memory controllers/channels in the SG2044 is 2 to 1, whereas it is 16 to 1 in the SG2042. Based on the results in this section, along with the memory bandwidth results from the STREAM benchmark in Figure \ref{fig:stream}, we conclude that these components become saturated in the SG2042 beyond a ratio of 4 to 1 whereas they are able to comfortably handle the maximum ratio of 2 to 1 in the SG2044.

We experimented with \emph{OMP\_PROC\_BIND} and \emph{OMP\_PLACES} to explore whether it was possible to obtain higher performance on the SG2044 by adopting different thread placement strategies. Different settings impacted performance, but consistently the best performance was obtained by either leaving these environment variables unset or with \emph{OMP\_PROC\_BIND} set to \emph{false}. This disables thread affinity and OpenMP threads are not bound to specific cores, providing the operating system flexibility to move threads between cores at runtime. We found this surprising, as for a memory bound workload we had assumed that pinning threads to cores that are distributed across the CPU would be beneficial to share the load between the memory controllers, however it seems that the OS did a better job at runtime.

\begin{table*}[h]
    \centering
    \caption{For each NPB pseudo application, the number of times faster a specific CPU is than the SG2044 at a given number of cores for class C.}
    \label{tab:pseudo_apps}
    \begin{tabular}{|c|cccc|cccc|cccc|}
    \hline
      \textbf{Number} & \multicolumn{4}{c|}{\textbf{BT benchmark}}  & \multicolumn{4}{c|}{\textbf{LU benchmark}} & \multicolumn{4}{c|}{\textbf{SP benchmark}}\\
      \textbf{cores} & SG2042 & EPYC & Skylake & ThunderX2 & SG2042 & EPYC & Skylake & ThunderX2 & SG2042 & EPYC & Skylake & ThunderX2\\ 
      \hline
      16 & 0.79 & 2.56 & 2.60 & 1.92 & 0.85 & 3.09 & 3.52 & 2.43 & 0.79 & 3.99 & 3.07 & 2.87\\
      26 & 0.66 & 2.35 & 1.95 & 1.77 & 0.88 & 2.80 & 2.77 & 2.29 & 0.57 & 3.56 & 1.99 & 2.05\\
      32 & 0.66 & 2.41 & - & 1.73 & 0.81 & 2.76 & - & 2.39 & 0.63 & 3.30 & - & 2.02\\
      64 & 0.45 & 1.90 & - & - & 0.69 & 2.05 & - & - & 0.48 & 2.05 & - & - \\
    \hline
    \end{tabular}
\end{table*}

\subsection{Embarrassingly Parallel (EP)}

\begin{figure}[htb]
    \centering
    {\includegraphics[width=1.0\columnwidth]{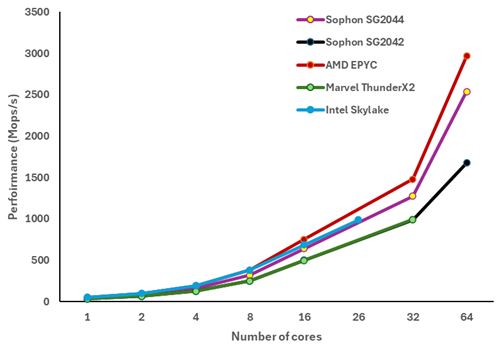}}
    \caption{EP benchmark performance (higher is better) parallelised via OpenMP.}
    \label{fig:ep}
\end{figure}

The Embarrassingly Parallel (EP) benchmark is compute bound, and results of this on our CPUs of interest are illustrated in Figure \ref{fig:ep}. When comparing performance against the SG2042, \cite{brown2023performance} observed that core for core there were two grouping, the Marvel ThunderX2 and Sophon SG2042, and the Intel Skylake and AMD EPYC. The SG2044 tracks performance of the Intel Skylake core-for-core very closely and once beyond 26 cores then follows a similar scaling trajectory to the AMD EPYC albeit at slightly lower performance in absolute terms. 

The compute performance delivered by the SG2044 is impressive and competitive against the other CPUs. Especially when one considers that the SG2044 is aimed more at workstation and server grade computing, rather than top-end high performance, and is likely to be cheaper than these other CPUs.

\subsection{Conjugate Gradient (CG)}

\begin{figure}[htb]
    \centering
    {\includegraphics[width=1.0\columnwidth]{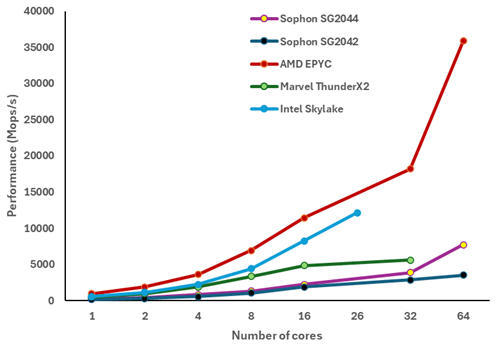}}
    \caption{CG benchmark performance (higher is better) parallelised via OpenMP.}
    \label{fig:cg}
\end{figure}

Figure \ref{fig:cg} illustrates the performance of the Conjugate Gradient (CG) benchmark kernel across our CPUs of interest. CG comprises irregular memory access and nearest neighbour communications, placing stress on the memory subsystem and potentially L2 and L3 shared cache. For smaller numbers of threads the SG2044 and SG2042 deliver similar performance, and it can be seen that the 2.2 times difference in performance observed in the multi-core comparison of Section \ref{sec:sg2042-compare} only starts building at 32 threads. Whilst, core for core, the Marvel ThunderX2 outperforms the SG2044, when running over the entire CPU 64 cores in the SG2044 outperforms 32 cores of the Arm CPU. 

Whilst the improved memory performance of the SG2044 compared with the SG2042 will be making a significant difference for this kernel, potentially the doubling of L2 cache, to 2 MB shared between groups of four cores, could also be having an impact due to the nearest neighbour communications. It should be noted that, unlike the other benchmarks, vectorisation was disable on the SG2044 for this kernel which is discussed in Section \ref{sec:compiler}.

\subsection{Fast Fourier Transform (FT)}

\begin{figure}[htb]
    \centering
    {\includegraphics[width=1.0\columnwidth]{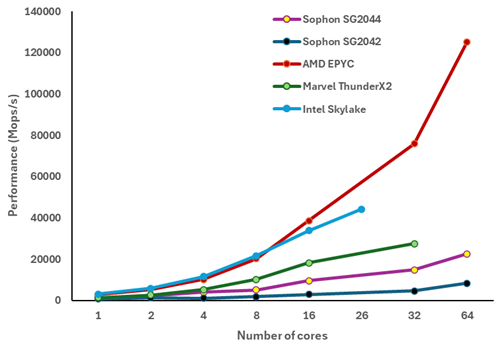}}
    \caption{FT benchmark performance (higher is better) parallelised via OpenMP.}
    \label{fig:ft}
\end{figure}

The fast Fourier Transform (FT) benchmark involves all-to-all communication between threads, and Figure \ref{fig:ft} reports the performance of this benchmark. It can be seen that the scaling behaviour of the SG2044 and SG2042 follow a similar trajectory, although offset in absolute terms. Whilst the SG2044 outperforms the SG2042 for this kernel, it still lags behind the other architectures.

\subsection{Pseudo Applications}

The BT, LU and SP pseudo applications are more complex benchmarks than the NPB kernels studied thus far. Table \ref{tab:pseudo_apps} reports the runtime of each architecture relative to the SG2044 at different numbers of cores for each pseudo application at class C, where a value of 1.0 represents an equal runtime. Broadly, it can be observed that as the number of cores increases the performance gap with the SG2042 widens. By contrast, in general, as the number of cores increases the SG2044 closes the performance gap with the other architectures. This is consistent with the results observed for the kernels and also Figure \ref{fig:stream}, where for the memory bound kernels (IS and MG) along with CG, the main benefit of the SG2044 compared to the SG2042 is that it can scale with the number of cores, closing the gap between this RISC-V CPU and other architectures where the SG2042 has plateaued. 

\section{Compiler differences and vectorisation}
\label{sec:compiler}

The system we are using to test the SG2044 runs the openEuler Linux operating system which ships GCC version 12.3.1. However, at the time of writing the latest release of GCC is version 15.2, and there have been significant improvements for RISC-V between these compiler versions. These include GCC v13.1 providing foundational support for vectorisation and GCC v14 providing full RVV v1.0 compatibility and improved RISC-V auto-vectorization. Hence for all experiments thus far we have used GCC v15.2 on the SG2044.

\begin{table}[htb]
    \centering
    \caption{Performance of NAS kernels (class C) running on a single core of the SG2044 with different compiler versions and with or without vectorisation.}
    \label{tab:sg_singlecore_compiler}
    \begin{tabular}{|c|c|c|c|}
    \hline
      \textbf{Benchmark}   & \makecell{\textbf{GCC v12.3.1} \\ \textbf{\textit{Mop/s}}} & \makecell{\textbf{GCC v15.2} \\ \textbf{vector enabled} \\ \textbf{\textit{Mop/s}}} & \makecell{\textbf{GCC v15.2} \\ \textbf{no vector} \\ \textbf{\textit{Mop/s}}} \\
     \hline
	  \textbf{IS} & 62.94 & 63.63 & 62.75 \\        
	  \textbf{MG} & 1373.31 & 1382.92 & 1300.27 \\
        \textbf{EP} & 40.56 & 40.76 & 40.75 \\
        \textbf{CG} & 210.06 & 81.19 & 217.53 \\
        \textbf{FT} & 887.43 & 1023.83 & 982.93 \\              
    \hline
    \end{tabular}
\end{table}

Table \ref{tab:sg_singlecore_compiler} reports performance on a single SG2044 C920v2 core for the benchmark kernels using GCC v12.3.1, and GCC v15.2 with and without vectorisation. It can be seen that GCC v15.2 always delivers higher performance than GCC v12.3.1 and generally vectorisation provides better performance. The EP, compute bound, benchmark surprised us as vectorisation delivers limited benefit. The exception is the CG benchmark, where the vectorised executable is around three times slower than the non-vectorised one. 

\begin{table}[htb]
    \centering
    \caption{Performance of NAS kernels (class C) running on all 64 cores of the SG2044 with different compiler versions and with or without vectorisation.}
    \label{tab:sg_multicore_compiler}
    \begin{tabular}{|c|c|c|c|}
    \hline
      \textbf{Benchmark}   & \makecell{\textbf{GCC v12.3.1} \\ \textbf{\textit{Mop/s}}} & \makecell{\textbf{GCC v15.2} \\ \textbf{vector enabled} \\ \textbf{\textit{Mop/s}}} & \makecell{\textbf{GCC v15.2} \\ \textbf{no vector} \\ \textbf{\textit{Mop/s}}} \\
     \hline
	  \textbf{IS} & 2255.72 & 3038.14 & 3024.63 \\        
	  \textbf{MG} & 32186.04 & 32457.83 & 31892.70 \\
        \textbf{EP} & 2529.91 & 2542.53 & 2538.38 \\
        \textbf{CG} & 7709.53 & 4463.18 & 7728.80 \\
        \textbf{FT} & 20796.20 & 22582.20 & 21282.00 \\              
    \hline
    \end{tabular}
\end{table}

Table \ref{tab:sg_multicore_compiler} explores multi-core performance, over all 64 cores of the SG2044, for these two compilers, and with and without vectorisation. With the exception of the CG benchmark, it can be seen that vectorisation is beneficial. The IS benchmark shows the greatest performance improvement for GCC v15.2 compared with v12.3.1, and the FT benchmark also benefits considerably but not by as much.

The CG kernel is the anomaly here where vectorisation significantly reduces performance. Throughout our experiments we used the default version of the code, but there are two alternative implementations of the matrix-vector multiplication within the \emph{cong\_grad} function unrolling the loop by two or eight times. We experimented with both of these and found that vectorised performance of the two-times unroll was 1.12 that of the default, and the eight-times unroll was 1.64. However, these still fell short of the non-vectorised performance. Running with \emph{perf} we noted that the vectorised executable suffers from around double the number of branch misses than the non-vectorised version, and the non-vectorised version is completing on average 0.54 instructions per cycle compared to 0.51 for the vectorised version. We also observed some performance reduction on the SpacemiT K1/M1 when vectorising the CG kernel, however this was marginal.

\section{Conclusions}
\label{sec:conclusions}

In this paper we have benchmarked SOPHGO's Sophon SG2044 against it's predecessor the SG2042 and a range of other RISC-V and non RISC-V CPUs. Enhancements present in the SG2044 addresses performance bottlenecks of the SG2042, but whilst there is some benefit for single core workloads (between 1.08 and 1.30 times faster) the ability for the SG2044 to continue performing as cores are increased, especially the memory subsystem, delivers significant performance (between 1.52 and 4.91) over and above the SG2042 for workloads running across all 64 cores.  

The two most important enhancements to the SG2044 for the HPC community is the improved memory subsystem and support for RISC-V vector extension (RVV) version 1.0. The ability to use a mainline compiler to target vectorisation is of significant benefit and the compute bounds kernel, EP, enjoyed the greatest benefit on a single C920v2 core of the SG2044 compared to the C920v1 in the SG2042. By contrast, improvements to the memory subsystem by SOPHGO engineers seem to have been focussed at the multi rather than single core level. Consequently, when compared against other architectures the SG2044 tends to perform less well for small numbers of cores and then catches up and becomes more competitive as the number of cores is increased. Typically in HPC one runs across all cores in the CPU, however it does mean that there is further for the SG2044 to go to catch the performance of these other architectures. This is likely, in part, due to SOPHGO's decision to continue using the same, albeit upgraded, C920 core in the SG2044 and enhance the subsystems around it.

Further exploration around the code properties within the CG kernel that reduce performance when vectorising would be very worthwhile. This would help to develop best practice for, and a better understanding of, the vectorisation of codes on the SG2044 and RVV in general, as well as acting as a case study of profiling codes on the SG2044. It would also be interesting to expand the number of benchmarks to include other HPC standard tests including HPCG and Linpack, as well as exploring the LLVM compiler which has supported RVV longer than GCC.

\begin{acks}
We thank SOPHGO for access to the SG2044 that was used for our experiments in this paper, and their engineers for their advice and insights about the CPU. This work was funded by the CONTINENTS EPSRC project grant number EP/Z531170/1 and HPC-R EPSRC project grant number EP/Z533701/1. This work used the ARCHER2 UK National Supercomputing Service (\url{https://www.archer2.ac.uk}). For the purposes of open access, the author has applied a CC BY public copyright license to any Author Accepted Manuscript version arising from this submission.
\end{acks}

\bibliographystyle{ACM-Reference-Format}
\bibliography{citations}


\begin{thebibliography}{15}


\ifx \showCODEN    \undefined \def \showCODEN     #1{\unskip}     \fi
\ifx \showISBNx    \undefined \def \showISBNx     #1{\unskip}     \fi
\ifx \showISBNxiii \undefined \def \showISBNxiii  #1{\unskip}     \fi
\ifx \showISSN     \undefined \def \showISSN      #1{\unskip}     \fi
\ifx \showLCCN     \undefined \def \showLCCN      #1{\unskip}     \fi
\ifx \shownote     \undefined \def \shownote      #1{#1}          \fi
\ifx \showarticletitle \undefined \def \showarticletitle #1{#1}   \fi
\ifx \showURL      \undefined \def \showURL       {\relax}        \fi
\providecommand\bibfield[2]{#2}
\providecommand\bibinfo[2]{#2}
\providecommand\natexlab[1]{#1}
\providecommand\showeprint[2][]{arXiv:#2}

\bibitem[bananapi(2024)]%
        {bananapi}
bananapi \bibinfo{year}{2024}\natexlab{}.
\newblock \bibinfo{booktitle}{\emph{Banana Pi BPI-F3 SpacemiT K1 RISC-V chip datasheet}}.
\newblock
\urldef\tempurl%
\url{https://docs.banana-pi.org/en/BPI-F3/SpacemiT_K1_datasheet}
\showURL{%
Retrieved Aug 9, 2025 from \tempurl}


\bibitem[Brown and Day(2025)]%
        {brown2025investigations}
\bibfield{author}{\bibinfo{person}{Nick Brown} {and} \bibinfo{person}{Christopher Day}.} \bibinfo{year}{2025}\natexlab{}.
\newblock \showarticletitle{Investigations of multi-socket high core count RISC-V for HPC workloads}. In \bibinfo{booktitle}{\emph{Proceedings of the 2025 International Conference on High Performance Computing in Asia-Pacific Region Workshops}}. \bibinfo{pages}{61--67}.
\newblock


\bibitem[Brown and Jamieson(2024)]%
        {brown2023performance}
\bibfield{author}{\bibinfo{person}{Nick Brown} {and} \bibinfo{person}{Maurice Jamieson}.} \bibinfo{year}{2024}\natexlab{}.
\newblock \showarticletitle{Performance characterisation of the 64-core SG2042 RISC-V CPU for HPC}. In \bibinfo{booktitle}{\emph{International Conference on High Performance Computing}}. Springer, \bibinfo{pages}{354--367}.
\newblock


\bibitem[Brown et~al\mbox{.}(2023)]%
        {brown2023risc}
\bibfield{author}{\bibinfo{person}{Nick Brown}, \bibinfo{person}{Maurice Jamieson}, \bibinfo{person}{Joseph Lee}, {and} \bibinfo{person}{Paul Wang}.} \bibinfo{year}{2023}\natexlab{}.
\newblock \showarticletitle{Is RISC-V ready for HPC prime-time: Evaluating the 64-core Sophon SG2042 RISC-V CPU}. In \bibinfo{booktitle}{\emph{Proceedings of the SC'23 Workshops of The International Conference on High Performance Computing, Network, Storage, and Analysis}}. \bibinfo{pages}{1566--1574}.
\newblock


\bibitem[Diehl et~al\mbox{.}(2024)]%
        {diehl2024preparing}
\bibfield{author}{\bibinfo{person}{Patrick Diehl}, \bibinfo{person}{Panagiotis Syskakis}, \bibinfo{person}{Gregor Dai{\ss}}, \bibinfo{person}{Steven~R Brandt}, \bibinfo{person}{Alireza Kheirkhahan}, \bibinfo{person}{Srinivas~Yadav Singanaboina}, \bibinfo{person}{Dominic Marcello}, \bibinfo{person}{Chris Taylor}, \bibinfo{person}{John Leidel}, {and} \bibinfo{person}{Hartmut Kaiser}.} \bibinfo{year}{2024}\natexlab{}.
\newblock \showarticletitle{Preparing for HPC on RISC-V: Examining Vectorization and Distributed Performance of an Astrophysics Application with HPX and Kokkos}. In \bibinfo{booktitle}{\emph{SC24-W: Workshops of the International Conference for High Performance Computing, Networking, Storage and Analysis}}. IEEE, \bibinfo{pages}{1656--1665}.
\newblock


\bibitem[Jin et~al\mbox{.}(1999)]%
        {jin1999openmp}
\bibfield{author}{\bibinfo{person}{Hao-Qiang Jin}, \bibinfo{person}{Michael Frumkin}, {and} \bibinfo{person}{Jerry Yan}.} \bibinfo{year}{1999}\natexlab{}.
\newblock \showarticletitle{The OpenMP implementation of NAS parallel benchmarks and its performance}.
\newblock  (\bibinfo{year}{1999}).
\newblock


\bibitem[Open chip community(2023)]%
        {c906}
Open chip community \bibinfo{year}{2023}\natexlab{}.
\newblock \bibinfo{booktitle}{\emph{Open XuanTie C906}}.
\newblock
\urldef\tempurl%
\url{https://xrvm.com/cpu-details?id=4056751997003636736}
\showURL{%
Retrieved Aug 16, 2023 from \tempurl}


\bibitem[rv-nvidia(2025)]%
        {rv-nvidia}
rv-nvidia \bibinfo{year}{2025}\natexlab{}.
\newblock \bibinfo{booktitle}{\emph{How NVIDIA Shipped One Billion RISC-V Cores In 2024}}.
\newblock
\urldef\tempurl%
\url{https://riscv.org/blog/2025/02/how-nvidia-shipped-one-billion-risc-v-cores-in-2024/}
\showURL{%
Retrieved Aug 9, 2025 from \tempurl}


\bibitem[Saphir et~al\mbox{.}(1997)]%
        {saphir1997new}
\bibfield{author}{\bibinfo{person}{William Saphir}, \bibinfo{person}{Rob~F Van~der Wijngaart}, \bibinfo{person}{Alex Woo}, {and} \bibinfo{person}{Maurice Yarrow}.} \bibinfo{year}{1997}\natexlab{}.
\newblock \showarticletitle{New Implementations and Results for the NAS Parallel Benchmarks 2.}. In \bibinfo{booktitle}{\emph{PPSC}}. Citeseer.
\newblock


\bibitem[sg2044(2024)]%
        {sg2044-announce}
sg2044 \bibinfo{year}{2024}\natexlab{}.
\newblock \bibinfo{booktitle}{\emph{SG2042 Empowering RISC-V in High-Performance Computing}}.
\newblock
\urldef\tempurl%
\url{https://github.com/RISCVtestbed/riscvtestbed.github.io/blob/main/assets/files/hpcasia24/hpc_asia_wang.pdf/}
\showURL{%
Retrieved Aug 9, 2025 from \tempurl}


\bibitem[sg2044-config(2025)]%
        {sg2044-config}
sg2044-config \bibinfo{year}{2025}\natexlab{}.
\newblock \bibinfo{booktitle}{\emph{Sophgo-doc: SG2044}}.
\newblock
\urldef\tempurl%
\url{https://github.com/sophgo/sophgo doc/blob/main/SG2044/HowTo/Configuraton%20Info%20in%20INI%20file.rst}
\showURL{%
Retrieved Aug 9, 2025 from \tempurl}


\bibitem[sg2044-evb-geekbench(2025)]%
        {sg2044-evb-geekbench}
sg2044-evb-geekbench \bibinfo{year}{2025}\natexlab{}.
\newblock \bibinfo{booktitle}{\emph{Geekbench: SOPHGO SOPHGO SG2044 EVB vs Milk-V Pioneer}}.
\newblock
\urldef\tempurl%
\url{https://browser.geekbench.com/v6/cpu/compare/10287883?baseline=8910646}
\showURL{%
Retrieved Aug 9, 2025 from \tempurl}


\bibitem[sg2044-geekbench(2025)]%
        {sg2044-geekbench}
sg2044-geekbench \bibinfo{year}{2025}\natexlab{}.
\newblock \bibinfo{booktitle}{\emph{Geekbench: SOPHGO SG2044 RISC-V vs Milk-V Pioneer}}.
\newblock
\urldef\tempurl%
\url{https://browser.geekbench.com/v6/cpu/compare/8661173?baseline=8910646}
\showURL{%
Retrieved Aug 9, 2025 from \tempurl}


\bibitem[SiFive U74-MC Core Complex Manual(2021)]%
        {u74}
SiFive U74-MC Core Complex Manual \bibinfo{year}{2021}\natexlab{}.
\newblock \bibinfo{booktitle}{\emph{SiFive U74-MC Core Complex Manual}}.
\newblock
\urldef\tempurl%
\url{https://starfivetech.com/uploads/u74mc_core_complex_manual_21G1.pdf}
\showURL{%
Retrieved Mar 17, 2024 from \tempurl}


\bibitem[spacemit(2024)]%
        {spacemit}
spacemit \bibinfo{year}{2024}\natexlab{}.
\newblock \bibinfo{booktitle}{\emph{SpacemiT K1 is an octa-core 64-bit RISC-V AI CPU}}.
\newblock
\urldef\tempurl%
\url{https://docs.banana-pi.org/en/BPI-F3/SpacemiT_K1}
\showURL{%
Retrieved Aug 9, 2025 from \tempurl}


\end{thebibliography}

\end{document}